\documentclass[aps,pre,twocolumn,amsmath]{revtex4}

\usepackage{graphicx} \usepackage{verbatim}
\usepackage[pdftex]{hyperref}
\hypersetup{colorlinks=true,linkcolor=blue,citecolor=blue,urlcolor=blue}

\begin{document}

\title{Static and dynamical properties of a hard-disk fluid confined to a narrow channel}

\author{M. J. Godfrey}
\author{M. A. Moore}
\affiliation{School of Physics and Astronomy, University of Manchester,
Manchester M13 9PL, UK}

\date{\today}

\begin{abstract}
The thermodynamic properties of disks moving in a channel sufficiently
narrow that they can collide only with their nearest neighbors can be
solved exactly by determining the eigenvalues and eigenfunctions of an
integral equation.  Using it we have determined the correlation length
$\xi$ of this system.  We have developed an approximate solution which
becomes exact in the high density limit.  It describes the system in
terms of defects in the regular zigzag arrangement of disks found in
the high-density limit.  The correlation length is then effectively
the spacing between the defects.  The time scales for defect creation
and annihilation are determined with the help of transition-state
theory, as is the diffusion coefficient of the defects, and these
results are found to be in good agreement with molecular dynamics
simulations.  On compressing the system with the
Lubachevsky--Stillinger procedure, jammed states are obtained whose
packing fractions $\phi_J$ are a function of the compression rate
$\gamma$.  We find a quantitative explanation of this dependence by
making use of the Kibble--Zurek hypothesis.  We have also determined
the point-to-set length scale $\xi_{PS}$ for this system.  At a
packing fraction $\phi$ close to its largest value
$\phi_{\text{max}}$, $\xi_{PS}$ has a simple power law divergence, $
\xi_{PS} \sim 1/(1-\phi/\phi_{\text{max}})$, while $\xi$ diverges much
faster, $\ln(\xi) \sim 1/(1-\phi/\phi_{\text{max}}) $.

\end{abstract}



\maketitle
\section{Introduction}
\label{Introduction}
Glasses and supercooled liquids have attracted a great deal of
attention from both experimentalists and simulators, but despite this
no totally satisfactory description of them is available.  There is a
suggestion that there is a connection between glassy behavior and
jammed states~\cite{LiuNagel}.  In this paper we shall examine a model
-- a system of hard disks confined to move in a narrow channel --
which is sufficiently simple that we can calculate analytically
quantities which in two and three dimensional systems have not yet
been satisfactorily studied despite extensive numerical efforts.
Bowles and colleagues \cite{Mahdi, Ivan, Ashwin} have studied this
model primarily by numerical methods and have elucidated many of its
features, including the numbers and properties of the jammed states
and the dynamics of the fluid states.  In this paper, we continue their
studies but our approach is primarily analytic.  This has the advantage
of providing physical insights as to what is going on.

\begin{figure}
\begin{center}
\includegraphics[width = 3.5in]{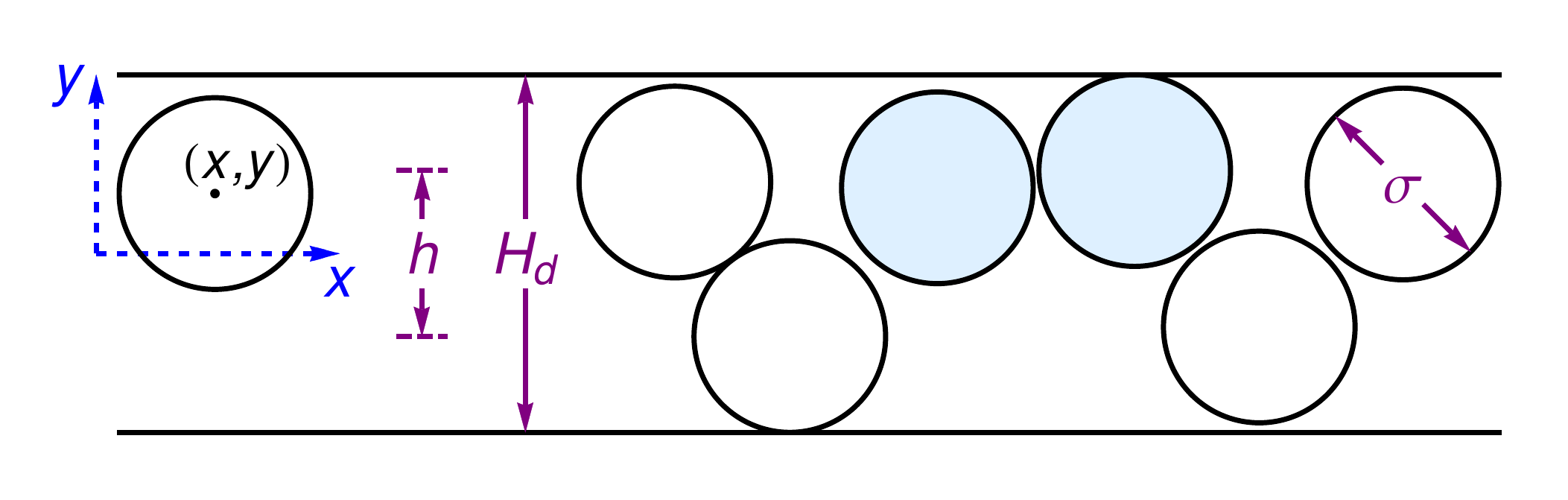}
\caption{(Color online) Geometry of the disks in a narrow
  channel.  The disks are of diameter $\sigma$ and the channel is of
  width $H_d$.  $h=H_d -\sigma$ is the width of the channel which is
  available to the centers of the disks.  The coordinates of the center
  of $i$th disk are $(x_i,y_i)$, where $y_i$ is measured from the
  centerline of the channel.  The blue-shaded disks are a defect in
  the zigzag arrangement of the disks that is favored at high
  density.}
\label{Ivanpicture}
\end{center}
\end{figure}

The model consists of $N$ disks of diameter $\sigma$, which move in a
narrow channel consisting of two impenetrable walls (lines) spaced by
a distance $H_d$ such that $1<H_d/\sigma< 1+\sqrt{3/4}$ (see
Fig.~\ref{Ivanpicture}).  The upper limit is imposed so that only
nearest-neighbor disk interactions can arise; also the disks cannot
pass each other, so their initial ordering is preserved for all times.
The disks and the walls are hard, so that configurations of the disks
where the centers approach by a distance less than $\sigma$ cannot
occur, and the center of each disk must be at a distance of at least
$\sigma/2$ from each wall.  It is useful to introduce the following
notation, which is also illustrated in Fig.~\ref{Ivanpicture}.  Let
the Cartesian coordinates of the center of disk~$i$ be denoted by
$(x_i,y_i)$, where the $x$-axis coincides with the center-line of the
channel.  A \emph{configuration} is a set of disk positions
$(x_i,y_i)$, $i =1,\,\ldots, N$, that is consistent with the
constraints of no overlap.  Let $h=H_d-\sigma$.  Then, because of the
hard walls, the allowed range of $y_i$ is $-h/2 \le y_i\le h/2$.  The
packing fraction or occupied volume is $\phi=N \pi \sigma^2/(4LH_d)$,
where $L$ is the length of the channel along the $x$-direction.  The
maximum possible value of $\phi$ will be called $\phi_{\text{max}}$.
It is given by
\begin{equation}
\phi_{\text{max}}=\frac{\pi \sigma^2}{4H_d\sqrt{\sigma^2-h^2}}.
\label{phimax}
\end{equation}
The numerical work described in this paper has been done for the case
when $h=\sqrt{3/4}\,\sigma$, for which $\phi_{\text{max}} \simeq
0.8418$.  A possible configuration of the disks is shown in
Fig.~\ref{Ivanpicture}, while the configuration associated with the
maximum possible packing fraction, $\phi_{\text{max}}$, is the zigzag
configuration shown in Fig.~\ref{twodefecttransitionstate}(a).  Note
that when $h =\sqrt{3/4}\,\sigma$, the centers of the disks form a
regular array of equilateral triangles in this, the most densely
packed state.

In Sec.~\ref{transfermatrix} we describe the transfer matrix formalism
that enables us to calculate exactly the equilibrium static properties
from the eigenvalues and eigenfunctions of an integral equation.  One
can determine the equation of state of the system from the largest
eigenvalue and its associated eigenfunction, and the correlation
length $\xi$ of the system is given by the logarithm of the ratio of
the largest and next-largest eigenvalues.  The transfer matrix
formalism gives few insights as to what is going on, so in
Sec.~\ref{theta} we discuss an analytical approximation which becomes
exact in the high density limit $\phi \to \phi_{\text{max}}$.  This
leads us to understand the nature of the order that is growing in the
system at high density.  This order is the zigzag arrangement of the
disks, which, for $\phi < \phi_{\text{max}}$, can be interrupted by
defects, like the blue-shaded disks in Fig.~\ref{Ivanpicture}
and also in Figs.~\ref{transitionstate} and
\ref{twodefecttransitionstate}.  It is shown that the correlation
length $\xi$ is a measure of the distance between the defects.  We
shall calculate $\theta $, which is the average concentration of
defects, as a function of the packing fraction $\phi$.

In Ref.~\cite{Ivan} it was found that, just as for two- and
three-dimensional hard sphere systems, there is a packing fraction
$\phi_d$ above which the dynamics becomes activated, and the activated
dynamics was studied as a function of the packing fraction $\phi$.  In
Sec.~\ref{dynamics} we shall show that this dynamics can be understood
analytically in the limit $\phi \to \phi_{\text{max}}$.  The same
approximation is fairly good over the entire range of $\phi$ between
$\phi_d$ and $\phi_{\text{max}}$.  The correlation length $\xi$ grows
rapidly for $\phi< \phi_d$ but does not diverge at $\phi_d$.
We are able to make analytical progress using the transition-state
approximation for those aspects of the dynamics associated with the
creation and annihilation of defects and their diffusion.  The
configurational entropy associated with the jammed states has been
calculated analytically \cite{Ashwin} and the same authors have used
the Lubachevsky--Stillinger algorithm \cite{LS} to determine how
$\phi_J$, the packing fraction at jamming, depends on the compression
rate.  In this paper we shall show that this dependence can be be
modeled by using the Kibble--Zurek~\cite{Kibble,Zurek} hypothesis.

In Sec.~\ref{point2set} we calculate the point-to-set length
$\xi_{PS}$.  It is much smaller than $\xi$ and has a quite different
dependence on $\phi$.  This might suggest that not all length scales
in glasses are fundamentally equivalent when they become large, but we
also point out that our system has some properties (most notably a
growing crystalline order) which are thought not to be of importance
in three dimensional glasses.

\section{Equilibrium Properties via the Transfer Matrix }
\label{transfermatrix}
In this section we set up the formalism by which the equation of state
and correlation length $\xi$ can be obtained, at least numerically,
from study of an integral equation.  We follow the procedure used in
Ref.~\cite{Kofke}.  The canonical partition function is
\begin{equation}
\exp(-\beta A_L)= \frac{1}{\Lambda^{Nd}} \prod_{i=1}^N\int dx_i \int_{-h/2}^{h/2}  dy_i \,I,
\end{equation}
where $(x_i,y_i)$ are the coordinates of the disk centers, with the
ordering $0<x_1<\ldots<x_N<L$, where $L$ is the length of the channel
available to the disk centers.  The integrand $I$ is 1 if the
configuration of $(x_i,y_i)$ is allowed but is zero if any two disks
overlap.  $d$ is the dimensionality of the channel, i.e.\ $d=2$.
$\Lambda=(2\pi\beta\hbar{}^2/m)^{1/2}$ is the thermal wavelength.  For
given values of the coordinates $y_i$ the system is isomorphic to a
mixture of one-dimensional hard rods of various lengths which allows
the integrations over the $x_i$ to be performed~\cite{Tonks}.  Then
\begin{align}
\exp(-\beta A_L)&=\nonumber\\
\frac{1}{\Lambda^{2N} N!} &\prod_{i=1}^N\int_{-h/2}^{h/2} dy_i\,
\bigl[L-\hbox{$\sum_{j=1}^{N-1} \sigma(y_j,y_{j+1})$}\bigr]^N\,I',
\nonumber
\end{align}
where $\sigma(y_i,y_{i+1})$ is the distance of closest approach of
neighboring disks $i$ and $i+1$ in the direction along the axis,
i.e.\ $[\sigma^2-(y_i-y_{i+1})^2]^{1/2}$.  The sum $\sum_{j=1}^{N-1}
\sigma(y_j,y_{j+1})$ is the total excluded volume of the ``hard
rods'', which must be smaller than $L$; this constraint is imposed
by the integrand $I'$, which can be either 0 or~1.

It is convenient to perform a Legendre transform of the Helmholtz free
energy $A_L$ by calculating the partition function
\begin{equation}
\exp (-\beta \Phi)=\beta f\int_0^{\infty}dL\, \exp(-\beta A_L) \exp (-\beta f L),
\label{legendre}
\end{equation}
where $f$ can be regarded as the force exerted by a piston at the end
of the channel; the longitudinal pressure $P$ is given by $P=f/h$, as
$h$ is the width of the channel accessible to the particle centers.  A
prefactor $\beta f$ has been introduced in (\ref{legendre}) to ensure
dimensional homogeneity; it is irrelevant to the thermodynamics.  The
integral over $L$ can now be performed, giving
\begin{align}
\exp(-\beta \Phi)={} &\frac{1}{(\beta f\Lambda^2)^N} \nonumber\\
&\times\prod_{i=1}^N \int_{-h/2}^{h/2}dy_i\, e^{-\beta f \sum_{j=1}^{N-1} \sigma(y_j,y_{j+1})}\,.
\label{Phidef}
\end{align}
The calculation of the potential $\Phi$ now manifestly involves only
nearest-neighbor interactions.  When $N$ is large, $\Phi$ is given
by~\cite{Kofke}
\begin{equation}
\beta\Phi \simeq N\ln\left(\beta f\Lambda^2/\lambda_1\right),
\end{equation}
where $\lambda_1$ (with the dimensions of length) is the largest
eigenvalue of the integral equation
\begin{equation}
\lambda_n\,u_n(y_1) =
\int_{-h/2}^{h/2}e^{-\beta f\sigma(y_1,y)}\,u_n(y)\,dy\,.
\label{inteqn}
\end{equation}
Equation (\ref{inteqn}) can be solved numerically by approximating the
integral by a sum, which leads to a symmetric matrix eigenvalue
problem~\cite{Kofke}.  When $\beta f\sigma$ is large, most of the
variation in the functions $u_n$ is concentrated near the walls of the
channel, so that it is helpful to make a transformation of the
variable $y$ before discretizing~\cite{Godfrey}.

All of the equilibrium properties of the system can, in principle, be
determined from the eigenvalues and eigenfunctions of~(\ref{inteqn}).
For example, the equation of state, i.e.\ the relation between $f$ and
the packing fraction $\phi$, can be found from~\cite{Kofke}
\begin{align}
  L/N = \frac1{\beta f} + \frac1{\lambda_1}&\int_{-h/2}^{h/2}\int_{-h/2}^{h/2}
  u_1(y_1)\,u_1(y_2)\nonumber\\
  &\times\sigma(y_1,y_2)\,e^{-\beta f\sigma(y_1,y_2)}\,dy_1\,dy_2\,,
\end{align}
which avoids the numerical differentiation that would be required to
calculate $L$ directly from $L=\partial \Phi/\partial f$.

\begin{figure}
\begin{center}
\includegraphics[width = 3.5in]{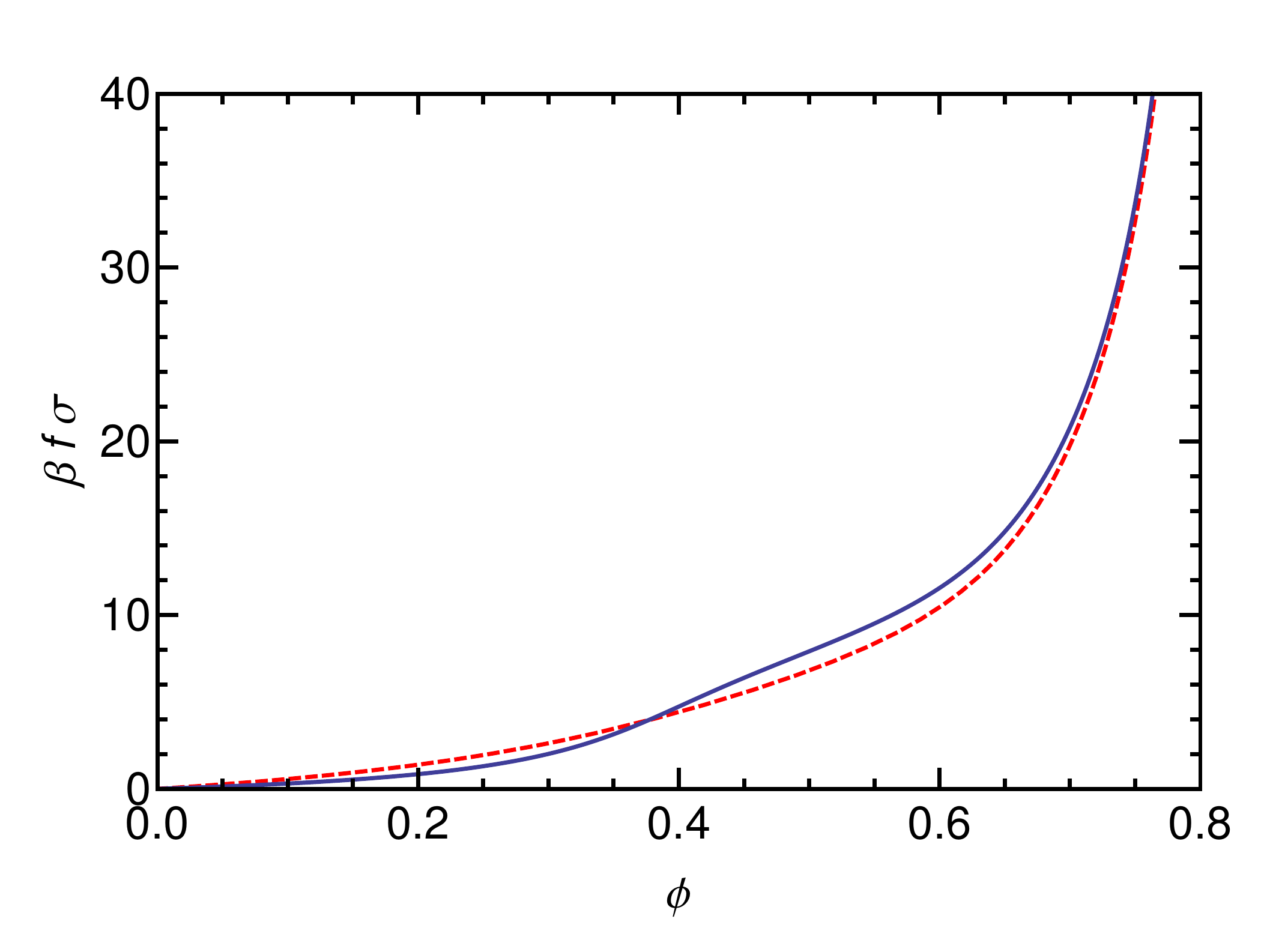}
\caption{(Color online) The equation of state of the narrow-channel
  system; that is, $\beta f \sigma$ versus the packing fraction
  $\phi$.  The solid line results are from the transfer matrix and are
  essentially exact.  Also shown (red dashed line) are the results
  from our analytical approximation Eqs.~(\ref{A},
  \ref{equationB}).  Note that at large $\phi$ there is excellent
  agreement between the two.}
\label{equationofstate}
\end{center}
\end{figure}

The plot of $\beta f \sigma $ versus $\phi$ is shown in
Fig.~\ref{equationofstate}.  We suspect that the ``shoulder'' that
appears near $\phi \simeq 0.5$ could be the remnant of the first
order transition which is seen in genuine two-dimensional
systems~\cite{Krauth}.  The pressure (or force) goes to infinity as the
density of the system approaches $\phi_{\text{max}}$.  The form of this
divergence will be discussed in Sec.~\ref{theta}.

\begin{figure}
\begin{center}
\includegraphics[width = 3.5in]{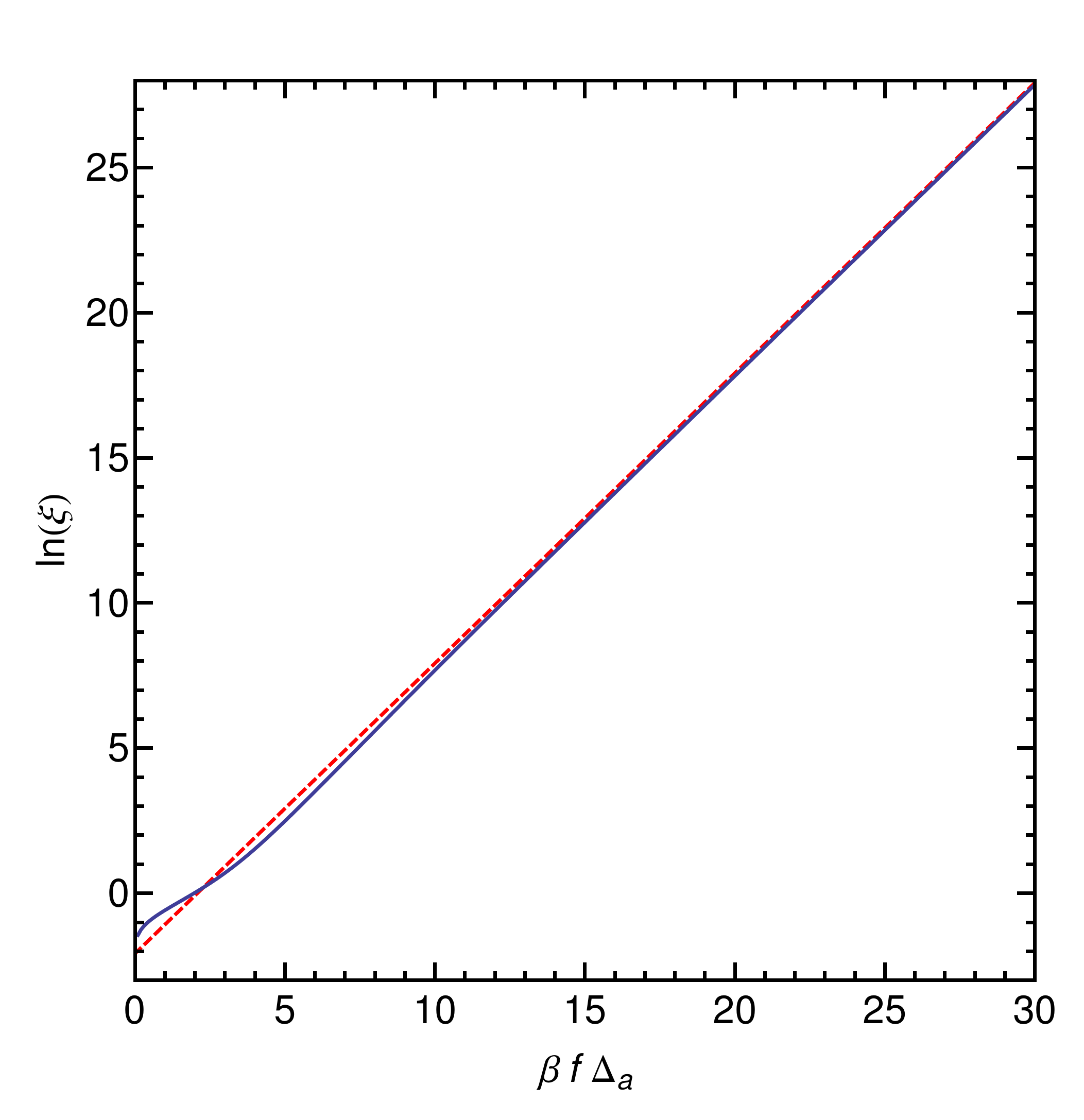}
\caption{(Color online) The logarithm of the correlation length $ \ln
  (\xi)$ versus $\beta f \Delta_a$ obtained from the transfer matrix
  (full line).  $\Delta_a$ is the extra length associated with a
  defect, which is discussed in Sec.~\ref{theta} at
  Eq.~(\ref{defDeltaa}).  It is given by
  $\Delta_a=\sigma-\sqrt{\sigma^2-h^2}$.  The red dashed line is the
  prediction of our analytical work, Eq.~(\ref{largexilimit}).}
\label{xi}
\end{center}
\end{figure}

The logarithm of the ratio of the two largest eigenvalues of the
integral equation gives a correlation length $\xi$, which is plotted
in Fig.~\ref{xi}.
The meaning of $\xi$ is that it is the number of disks that typically
form the zigzag pattern seen in the defect-free regions of
Fig.~\ref{Ivanpicture}.  More precisely, it measures the decay of the
correlation between the $y$-coordinates of well-separated disks $i$
and~$i+s$,
\begin{equation}
|\langle y_i\, y_{i+s} \rangle| \sim \exp(-s/\xi)\,,
\label{xidef}
\end{equation}
where $s\gg1$.  Notice that $\xi$ is dimensionless, because $s$ is an
integer; of course, if one knows the packing fraction, one can convert
$\xi$ into a distance by multiplying by the average separation of
neighboring disks.  Fig.~\ref{xi} shows that as $\phi \to \phi_{\text{max}}$,
$\xi$ grows rapidly, reflecting the fact that $\xi$ is essentially the
spacing between the defects, which becomes very large in this limit.
This observation will be made quantitative in Sec.~\ref{theta}.

\begin{figure}
\begin{center}
\includegraphics[width = 3.5in]{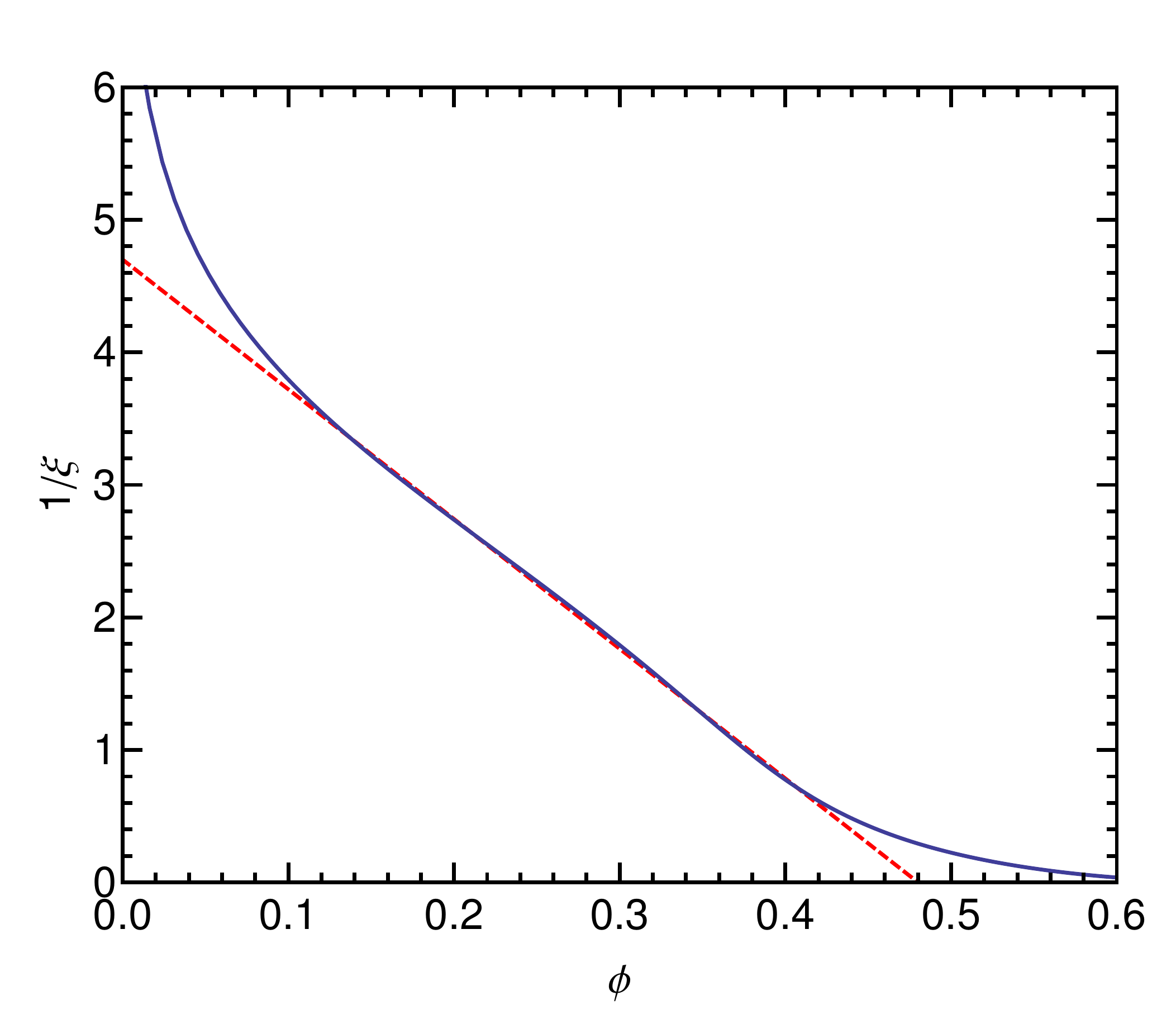}
\caption{(Color online) The reciprocal correlation length $1/\xi$
  versus the packing fraction $\phi$ (blue solid curve).  The red
  dashed line is a straight-line fit to highlight the rapid growth of
  $\xi$ as $\phi \to \phi_d \simeq 0.48$.  There is no true divergence
  of $\xi$ at the dynamical transition packing fraction $\phi_d$ and
  the ``transition'' is rounded-off.}
\label{inversexi}
\end{center}
\end{figure}

The behavior of $\xi$ at smaller packing fractions is also of
interest.  In Fig.~\ref{inversexi} we have plotted the reciprocal of
$\xi$ versus the packing fraction.  The fitted straight line
extrapolates to a value $\phi_d \simeq 0.48$.  This packing fraction
was identified in Ref.~\cite{Ivan} as the density above which the
dynamics becomes activated (see Fig.~\ref{Ivanfig}).  Similar behavior
arises for hard spheres in three dimensions at $\phi=\phi_d \simeq
0.58$ and is attributed there to the onset of caging.  In the
mode-coupling approximation this is accompanied by a diverging length
scale, but in a better approximation this length scale would be
expected to remain finite~\cite{Kobdyn}.  The same features seem to be
present in this system of disks in a channel.  The data in
Fig.~\ref{inversexi} shows that between $\phi=0.1$ up to $\phi=0.4$
the growth of $\xi$ is approximated well by
\begin{equation}
\xi \sim \frac{a}{1-\phi/\phi_d}.
\label{mctdivergence}
\end{equation}
Fig.~\ref{inversexi} shows that there is not a true divergence of
$\xi$ at $\phi_d$: the apparent divergence is rounded off and at
$\phi=\phi_d$, $\xi$ is approximately~4.  This just means that at this
density the growth of zigzag order has grown to involve four adjacent
disks, so that the disks will typically no longer be able to bounce
from one wall to another unless they first push other disks away.  In
other words, above $\phi_d$ the disks are caged.  Escape from the cage
requires the collective motion of several disks, a process which will
be studied in more detail in Sec.~\ref{dynamics}.  We believe also
that the onset of caging behavior is also ultimately responsible for
the strong fluid to fragile fluid crossover behavior described in
Ref.~\cite{Mahdi}. For $\phi > \phi_d$ the dynamics is fragile, as
the caging causes relaxation times to increase faster than would be
expected on a simple Arrhenius picture.

If the analogy with hard spheres near $\phi_d \simeq 0.58$ is
appropriate then a suitable autocorrelation function for studying
glassy behavior in our system would seem to be
\begin{equation}
A(t)=\frac{\langle y_i(0) y_i(t) \rangle}{\langle y_i^2 \rangle}.
\label{auto}
\end{equation}
$A(t)$ will equal unity at $t=0$ and, as $\phi \to \phi_d$, one would
expect a plateau to develop for larger times.  After a time
$\tau_{\alpha}$, the ``alpha'' relaxation time, $A(t)$ starts to decay
to zero as the disks escape their cages and $\langle y_i(t)\rangle$
approaches zero.  The study of this autocorrelation function will be
published separately~\cite{GodfreyB}.

Several groups have extracted a dynamical length scale from the four
point dynamical susceptibility of bulk colloidal matter, or via
simulations of it for hard spheres and disks~\cite{Keys, Narumi,
  Brambilla}.  In these studies the length scale relates to the
average number of particles that are cooperating in a dynamic
heterogeneous event.  We suspect that this dynamical length therefore
might relate to our length scale $\xi$, which is also a measure of the
number of particles which move cooperatively, but see
Ref.~\cite{Kobdyn}.

\section{Results for large densities}
\label{theta}
Using the integral equation (\ref{inteqn}) to solve for the
equilibrium properties gives little insight as to what is going on,
and it becomes increasingly difficult as $\phi \to \phi_{\text{max}}$.
Analytical progress is, however, possible in that limit, because every
disk is typically found within a small distance of order $1/(\beta f)$
from a wall.  The disks form the zigzag pattern shown in
Fig.~\ref{Ivanpicture} with relatively few defects, where a defect is
a pair of disks (like those shaded in Fig.~\ref{Ivanpicture}) that lie
close to the same wall of the channel.  We begin by calculating
$\theta$, the concentration of these defects present in the system at
equilibrium.

When $\beta f$ is large, the largest contributions to the partition
function (\ref{Phidef}) come from the neighborhoods of jammed states
in which every disk is in contact with its two neighbors and a wall of
the channel.  Each jammed state is a local minimum of the excluded
volume $\sum_j\sigma(y_j,y_{j+1})$, which is the reason its
neighborhood makes a relatively large contribution to the partition
function.  We calculate these contributions below.

It is convenient to introduce new integration variables to parametrize
a configuration in the neighborhood a jammed state.  We define $z_i$ to be the
distance of disk $i$ from its confining wall at $y=\pm h/2$.

For neighboring disks $1$ and $2$ on opposite sides of the channel,
the contribution to the excluded volume is
\begin{align}
\sigma(1,2)&=\sqrt{\sigma^2-(h-z_1-z_2)^2} \nonumber\\
&\simeq \sqrt{\sigma^2-h^2}+\frac{h}{\sqrt{\sigma^2 -h^2}} (z_1+z_2).
\label{distances}
\end{align}
Neighboring disks on the same side of the channel make a contribution
$\sigma(1,2) \simeq \sigma +\text{O}\left[(z_1-z_2)^2/
  \sigma\right]$; in this case, there is no term linear in $z_1$ or
$z_2$.

The jammed states can be more fully specified by stating the number
and arrangement of defects within them.  We suppose that there are $M$
defects in a particular jammed state.  Any one of the disks in this
state will either have both of its neighbors on the opposite side of
the channel if it forms part of the zigzag pattern, or it will have
one neighbor on the same side of the channel and the other on the
opposite side if it is part of a defect.  (Configurations in which
three or more disks lie adjacent at the same wall need not be
considered, as they are unstable under compression: there is no
barrier to moving the central disk to the opposite side of the
channel.)  We can renumber the disks $k=1$ to $2M$ for those belonging
to defect pairs and $k=2M+1$ to $N$ for those which do not.  With this
renumbering, and in terms of the new variables $z_k$, the excluded
volume can be written
\begin{align}
\sum_{i=1}^{N-1} \sigma(y_i,y_{i+1})\simeq{}&(N-M)\sqrt{\sigma^2-h^2}+M \sigma\nonumber\\
&+\sum_{k=1}^{2M} \frac{h z_k}{\sqrt{\sigma^2-h^2}}+\sum_{k=2M+1}^N \frac{2hz_k}{\sqrt{\sigma^2-h^2}},
\label{linearlength}
\end{align}
to first order in the variables~$z_k$.  We insert this expression into
Eq.~(\ref{Phidef}) and integrate from $z_k=0$ to $\infty\,$; formally,
$z_k$ should always be smaller than $h$, but when $\beta f$ is large
the error due to extending the range of integration is exponentially
small.  The resulting contribution to (\ref{Phidef}) is
\begin{widetext}
\begin{equation}
\frac{1}{(\beta f\Lambda^2)^N}
\left(\frac{\sqrt{\sigma^2-h^2}}{\beta f h}\right)^{2M}
\left(\frac{\sqrt{\sigma^2-h^2}}{2 \beta f h}\right)^{N-2M}
e^{-\beta f[(N-M) \sqrt{\sigma^2-h^2}+M\sigma]},
\label{partpart}
\end{equation}
\end{widetext}
which depends only on $M$, and not on the detailed arrangement of the
defects.  To obtain the total contribution from all states with $M$
defects, we must multiply (\ref{partpart}) by the number of these
arrangements, which is approximately
\begin{equation}
W_M=\frac{(N-M)!}{M!\,(N-2M)!}\,.
\label{combo}
\end{equation}
The combinatoric factor $W_M$ has a simple explanation.  Each of the
$M$ defects consists of a pair of neighboring disks.  This accounts
for $2M$ disks; the remaining $N-2M$ ``free'' disks are not part of
any defect.  The defect configurations can be regarded as arrangements
of $M$ defect-pairs and $N-2M$ free disks, in which adjacent
\emph{objects} (defect-pairs or free disks) occur alternately on
opposite sides of the channel.  The number of arrangements of $N-M$
objects, $M$ of one kind and $N-2M$ of another, is the factor $W_M$
given in Eq.~(\ref{combo}).

[The preceding argument for $W_M$ ignores the facts that, in a
  rectangular channel, the first and last disks must, for stability,
  be \emph{free} disks, and that, for any given $M$, there are two
  possible arrangements for this pair.  The remaining $N-2-2M$ free
  disks and $M$ defect-pairs can be permuted arbitrarily, which leads
  to $2\times(N-2-M)!/[(N-2-2M)!\,M!]$ as the correct combinatoric
  factor.  The difference compared with (\ref{combo}) is unimportant
  in the application below, in which we use the thermodynamic limit of
  $\ln W_M$.  It may be noted that we have also ignored the special
  nature of the first and last disks in the right-hand side of
  Eq.~(\ref{linearlength}).  Treating these correctly would change the
  exponents in (\ref{partpart}) by $\pm2$, which again is unimportant
  in the thermodynamic limit.]

\begin{figure}
\begin{center}
\includegraphics[width = 3.5in]{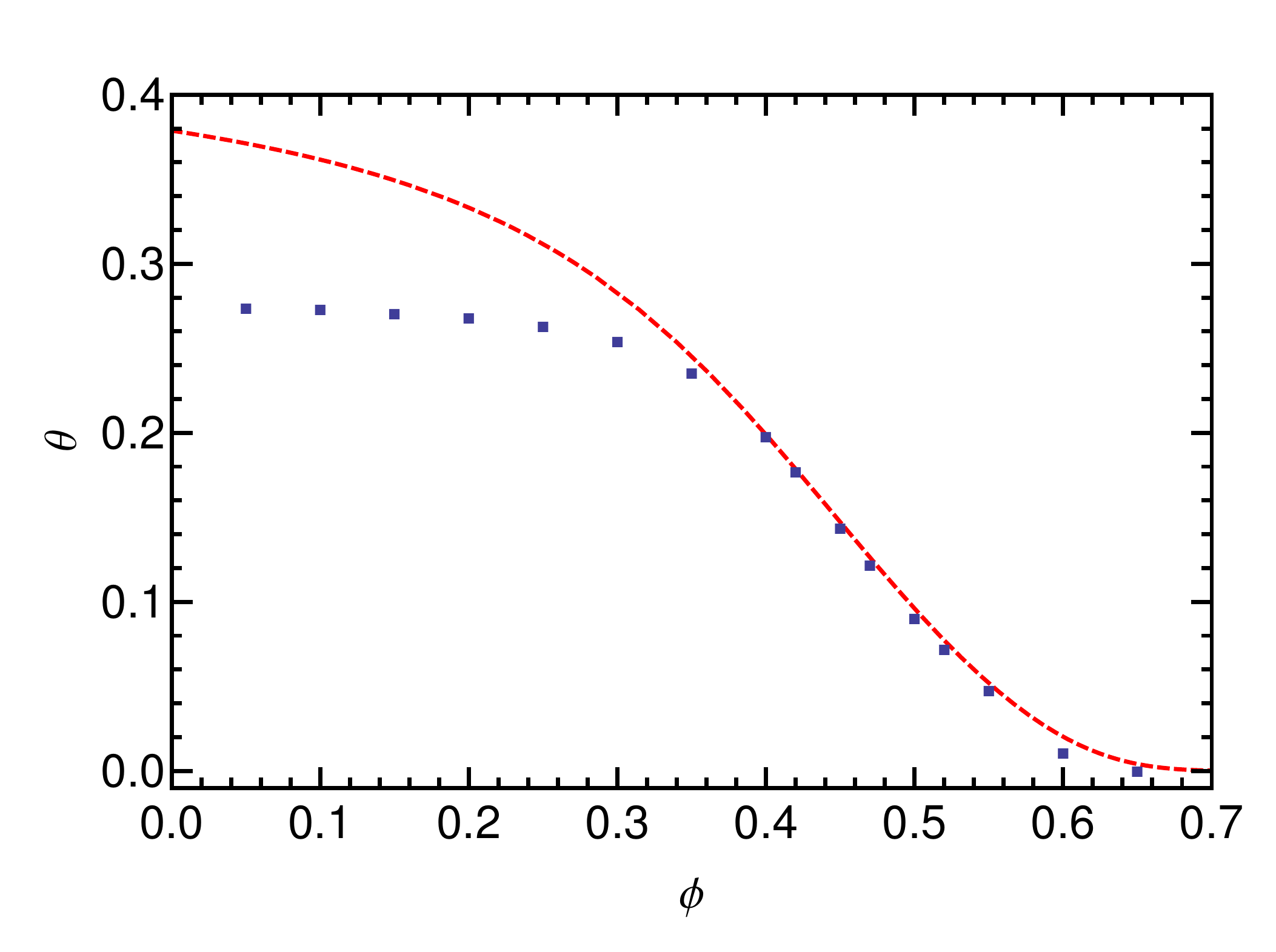}
\caption{(Color online) Plot of $\theta =\langle M\rangle/N$, the
  average density of defects, against $\phi$.  The data points are
  from the simulations in \cite{Ivan}, while the (red) dashed line is
  from solving Eqs.~(\ref{A}) and~(\ref{equationB}).  This analytical
  solution is expected to be an increasingly good approximation
  as $\phi$ approaches~$\phi_{\text{max}}$.}
\label{thetaversusphi}
\end{center}
\end{figure}

After combining the results of Eqs.~(\ref{partpart}) and
(\ref{combo}) we find
\begin{widetext}
\begin{equation}
\exp(-\beta \Phi)=\frac{1}{(\beta f\Lambda^2)^N}\sum_M W_M\, e^{-\beta f[(N-M) \sqrt{\sigma^2-h^2}+M\sigma]} \left(\frac{\sqrt{\sigma^2-h^2}}{\beta f h}\right)^N\frac1{2^{N-2M}}\,.
\end{equation}
In the thermodynamic limit we can convert the sum over $M$ to an
integral over $\theta$, where $M =\theta N$, and write
\begin{equation}
\exp(-\beta \Phi)= N\int d\theta\, \exp[ -\beta  \Phi^*(N, \beta f, \theta)]\,.
\label{Nlargeint}
\end{equation}
The effective free energy $\Phi^*$ is given by
\begin{align}
\beta \Phi^*(N, \beta f,\theta)=
-N \biggl\{&(1-\theta)\ln(1-\theta)-\theta \ln \theta
-(1-2\theta)\ln(1-2 \theta) \nonumber\\
&-\beta f\bigl[(1-\theta)\sqrt{\sigma^2-h^2}+\theta \sigma\bigr]
 +\ln \frac{\sqrt{\sigma^2-h^2}}{h} -2 \ln (\beta f \Lambda) -(1-2 \theta)\ln 2\biggr\},
\end{align}
\end{widetext}
in which we have used Stirling's approximation for the logarithms of
factorials.  For large $N$ the integral in Eq.~(\ref{Nlargeint}) can be
done by steepest descents by finding the solution of
$\partial\Phi^*(N, \beta f, \theta)/\partial \theta=0$.  This yields
an equation for the equilibrium value of the defect density $\theta$,
\begin{equation}
\frac{\theta (1-\theta)}{(1-2\theta)^2}=4 \exp[\beta f(\sqrt{\sigma^2-h^2}-\sigma)].
\label{A}
\end{equation}
From the relation $L=\partial \Phi^*/\partial f$ we can obtain the
equilibrium length of the system,
\begin{equation}
L=N[(1-\theta)\sqrt{\sigma^2-h^2}+\theta \sigma] +\frac{2 N}{\beta f}\,.
\label{B}
\end{equation}
This can be rearranged to give the approximate equation of state
\begin{equation}
\beta f= \frac{2 N}{L-N[(1-\theta)\sqrt{\sigma^2-h^2}+\theta \sigma]}.
\label{equationB}
\end{equation}
From Eqs.~(\ref{A}, \ref{equationB}) we can calculate $\beta f \sigma$
and $\theta$ in terms of $\phi$, as shown in
Figs.~\ref{equationofstate} and~\ref{thetaversusphi}.
In the high density limit, the agreement is excellent as would be
expected, but what is more surprising is that the agreement is fairly
good down to quite low values of the packing fraction, $\phi\simeq
\phi_d$.

In the limit when $\beta f \sigma $ is large, $\theta$ is small, and
Eq.~(\ref{A}) simplifies to
\begin{equation}
\theta\simeq 4 \exp[-\beta f(\sigma-\sqrt{\sigma^2-h^2})].
\label{Asimp}
\end{equation}
The exponential can be understood as an ordinary Boltzmann factor.
The extra length $\Delta_a$ involved in inserting a single defect into
the system over the length in the state of maximum density is
\begin{equation}
 \Delta_a=\sigma -\sqrt{\sigma^2-h^2}\,.
\label{defDeltaa}
\end{equation}
The work done increasing the length against the applied force is
$\Delta E=f \Delta_a$, so the exponential in Eq.~(\ref{Asimp}) is just
the usual Boltzmann expression $\exp(-\beta \Delta E)$.

The equation of state, Eq.~(\ref{equationB}), can be simplified in the
limit of large $\beta f \sigma$, when $\theta \to 0$, giving
\begin{equation}
\beta f \simeq\frac{2N}{L(1-\phi/\phi_{\rm{max}})},
\label{SWlimit}
\end{equation}
which is consistent with the general results of Salsburg and Wood
\cite{SW} for the limit $\phi \to \phi_{\text{max}}$.

Intuitively, one would expect there to be a relation between the
density of defects $\theta$ and the correlation length $\xi$, with
$\theta \sim 1/\xi\,$: the correlation length should be comparable
with the spacing between the defects.
\begin{figure}
\begin{center}
\includegraphics[width = 3.5in]{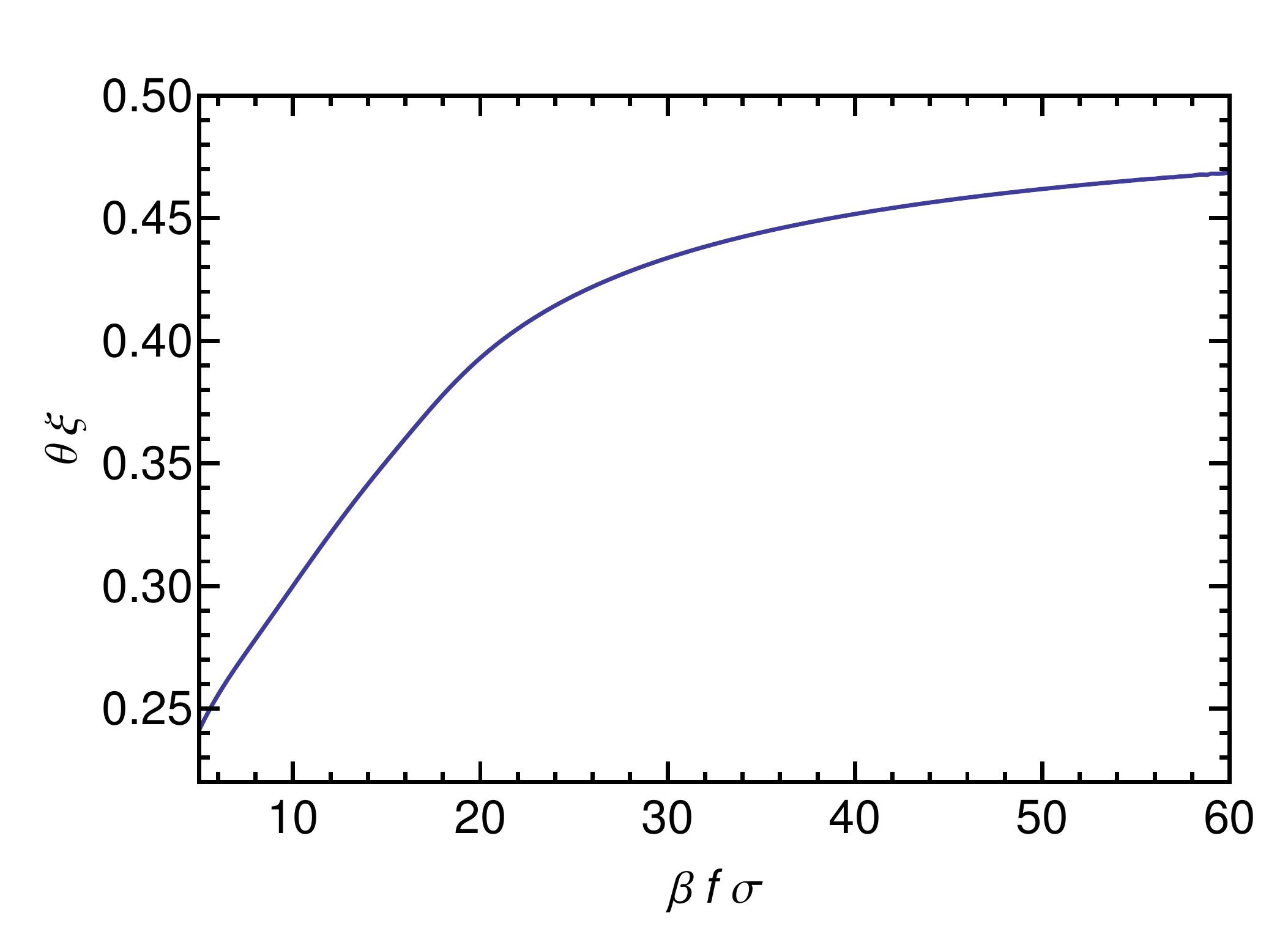}
\caption{(Color online) $\theta \xi$ versus $\beta f \sigma$.  Here
  $\theta$ was obtained from the solution of Eqs.~(\ref{A},
  \ref{equationB}) and $\xi$ from the transfer-matrix solution.  As
  $\beta f \sigma$ becomes large, Eq.~(\ref{xitheta}) predicts that
  $\theta \xi \to 1/2$. }
\label{thetaxi}
\end{center}
\end{figure}
Fig.~\ref{thetaxi} bears this out and further suggests that as $\phi
\to \phi_{\text{max}}$ the product $\theta \xi\to 1/2$.  This result
can be understood as follows.  In the limit $\phi \to
\phi_{\text{max}}$, the defects are very dilute and are almost
independent of one another.  The probability $P_k(r)$ that there will
be $k$ defects between disks $i$ and $i+r$ should therefore follow a
Poisson distribution,
\begin{equation}
P_k(r)=\frac{1}{k!} (\theta r)^k \exp(-\theta r).
\label{Poisson}
\end{equation}
In the high-density limit, the disks are pressed tightly against the
walls, so each $y_i$ is approximately $\pm h/2$.  The probability that
$y_i\,y_{i+r} \simeq + (h/2)^2$ is then the probability that there is an
even number of defects between $i$ and $i+r$, which is equal to the
sum
\begin{equation*}
P_{\text{even}}=\sum_{k=0}^{\infty} P_{2k}(r)=[1+e^{-2 \theta r}]/2\,,
\end{equation*}
while the probability that $y_i y_{i+r} \simeq -(h/2)^2$ is the
probability that there is an odd number of defects between $i$ and
$i+r$, which is the sum
\begin{equation*}
P_{\text{odd}}=\sum_{k=0}^{\infty}P_{2k+1}(r)=[1-e^{-2
    \theta r}]/2\,.
\end{equation*}
From these,
\begin{align}
\langle y_i y_{i+r}\rangle &\simeq(h/2)^2 \left(P_{\text{even}} +(-1)P_{\text{odd}}\right) \nonumber\\
&=(h/2)^2 \exp(-2 \theta r) \nonumber\\
&\equiv (h/2)^2 \exp(-r/\xi).
\label{xitheta}
\end{align}
Thus $2\theta=1/\xi$, or $\theta \xi=1/2$ at high density.  The form
of $\xi$ as $\phi \to \phi_{\text{max}}$ is therefore
\begin{equation}
\xi \simeq \hbox{$\frac18$}\exp(\beta f \Delta_a).
\label{largexilimit}
\end{equation}
This result is clearly consistent with the numerical results shown in
Fig.~\ref{xi}.  The argument of the exponential can also be understood
by calculating the defect free energy $\delta F \simeq f \Delta_a-k_B
T \ln \xi$.  The first term is the energy cost of creating the defect;
the second is the reduction in its free energy by the entropy of
placing it at any of $\xi$ positions.  Equating the defect free energy
to zero gives the exponential in Eq.~(\ref{largexilimit}).

While the system of disks in a narrow channel has been studied for the
insight it could provide on glass behavior in three dimensions, there
is one striking difference between it and typical three dimensional
glasses.  It is that we {\it understand} the origin in the narrow
channel system of its growing static length scale $\xi\,$: it
quantifies the growth of the zigzag order as $\phi \to
\phi_{\text{max}}$.  This growth would also be expected to be visible
in the structure factor $S(q_x, q_y)$ as Bragg-like peaks at the
wavevectors corresponding to that of the zigzag pattern (i.e.,
multiples of $q_x= 2\pi/\sqrt{\sigma^2-h^2}$ and $q_y=2\pi/h$) which
would grow as $\phi \to \phi_{\text{max}}$.  This feature of the
narrow-channel system reflects the fact that the most dense state has
crystalline order, and at lower densities the growing length scale
$\xi$ is a measure of the extent of the short-range crystalline order.
The crystalline order is broken by the (topological!)\ defects and the
correlation length $\xi$ is basically the spacing between the defects.
In three dimensions it is found that the structure factor hardly
alters at densities close to $\phi_d$, but it is sometimes suggested
that glass behavior might be associated with changes in more subtle
correlations; see e.g.\ \cite{Royall,Tanaka}.  We suspect that higher
correlation functions will describe the onset of caging and must
therefore contain a growing length scale.  But these correlation
functions have yet to be identified, and they may turn out to depend
on the details of the intermolecular potential.

\section{Dynamics}
\label{dynamics}
Glass behavior is largely a dynamic phenomenon and in this section we
shall analyse a few aspects of the dynamics of our system which are
sufficiently simple to permit an analytical treatment.

\begin{figure}
\begin{center}
\includegraphics[width = 3.5in]{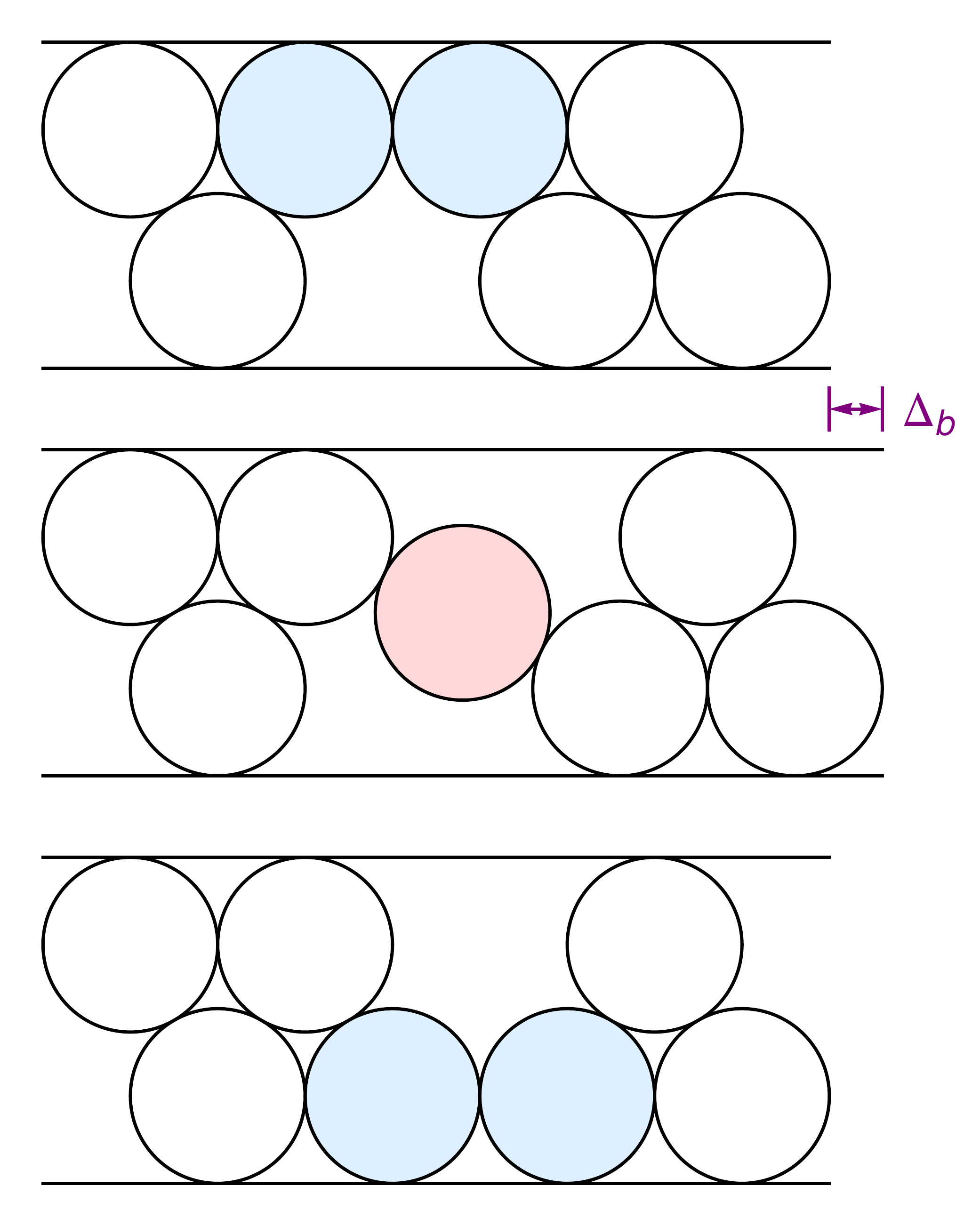}
\caption{(Color online) The transition state for motion of a defect.
  In the top diagram, the two blue-shaded disks are a defect in the
  zigzag arrangement of disks.  The defect can move when one disk
  crosses the channel by squeezing between its neighbors: the system
  passes through the transition state shown in the middle diagram to
  reach the defect state shown in the bottom diagram.  In the top
  diagram the defect involves disks $3$ and $4$; in the bottom diagram
  the defect involves disks $4$ and $5$, when the disks are numbered
  from the left.  The net motion of the defect is to the right, and
  $\Delta_b$ is the extra length needed to allow this motion.}
\label{transitionstate}
\end{center}
\end{figure}

We begin by calculating the typical time it takes for a defect to hop
to its neighboring site.  This was studied in Ref.~\cite{Ivan} by
means of molecular dynamics, but we shall use transition-state
theory~\cite{Barnett}, which works best when the motion is inhibited
and the transition rate is small.  The transition state is the state
through which the system has to squeeze during the course of a
transition: Fig.~\ref{transitionstate} shows the transition state that
our system of hard disks must pass through in order for a defect to
move. When the dynamics is not dominated by just the transition state,
the more general approach of studying the complete landscape as in
Ref.~\cite{Hunter} might be useful.

\begin{figure}
\begin{center}
\includegraphics[width = 3.5in]{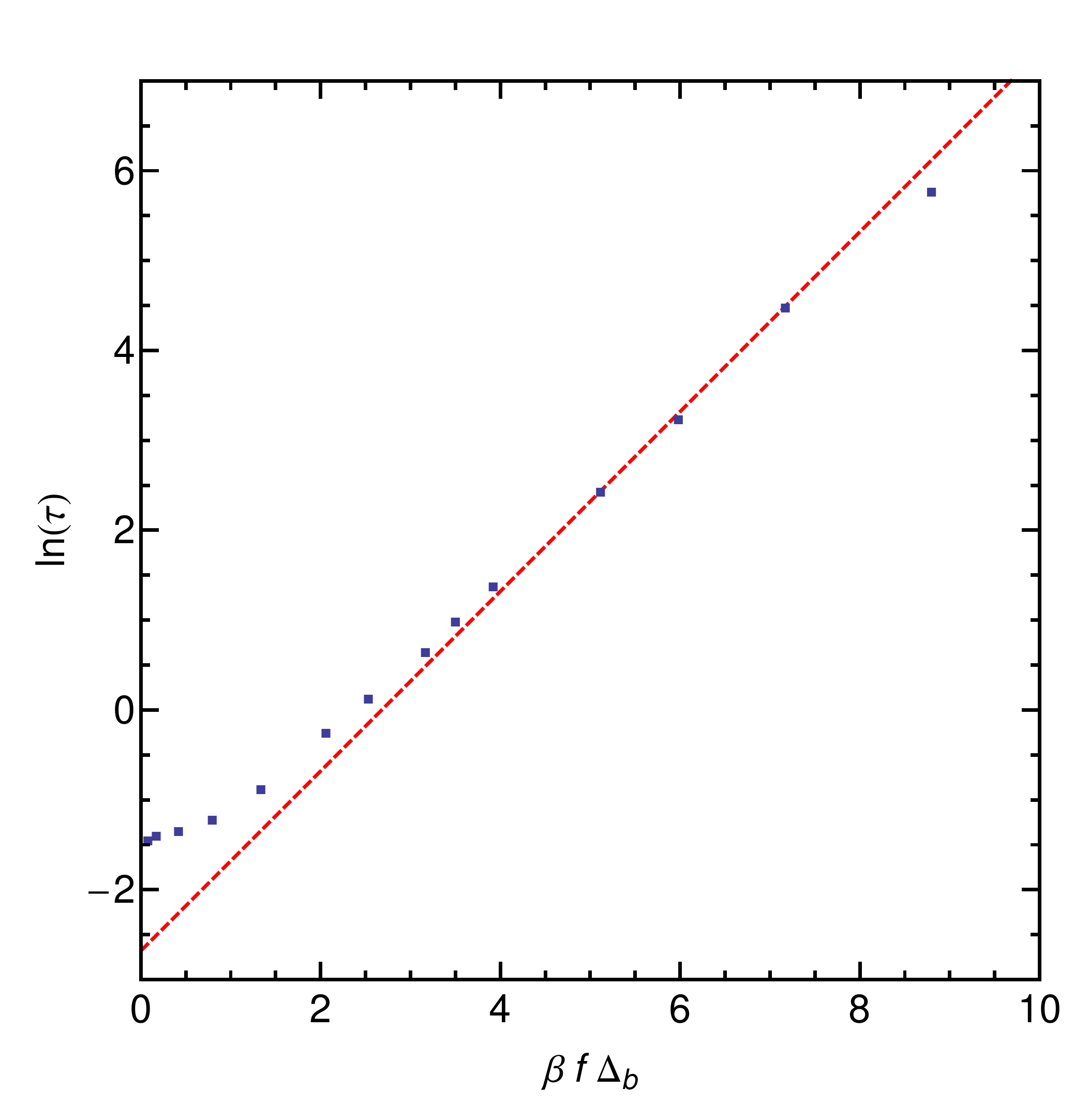}
\caption{(Color online) $\ln (\tau)$, where $\tau$ is the time scale
  on which defects move, plotted against $\beta f \Delta_b$.
  $\Delta_b$ is defined in Eq.~(\ref{saddle}) and is the extra length
  associated with the transition state through which the system must
  pass to allow a defect to move.  The red dashed line is the
  prediction of our analytical approximation
  Eq.~(\ref{rateofhopping}), which would be expected to be exact only
  for large $\beta f \Delta_b$.  Data points are from
  Ref.~\cite{Ivan}.}
\label{Ivanfig}
\end{center}
\end{figure}

At the transition state, the extra length of the system $\Delta_b$
over the length which just contains the defect is
\begin{equation}
\Delta_b= \sqrt{4 \sigma^2-h^2}-\sigma -\sqrt{\sigma^2-h^2}.
\label{saddle}
\end{equation}
The transition rate associated with this saddle is
\begin{equation}
1/\tau=1/\tau_0 \exp(-\beta f \Delta_b),
\label{rateofhopping}
\end{equation}
where $\tau_0$ is of the order of a disk collision time.  In
Fig.~\ref{Ivanfig} we have plotted data on $\tau$ from
Ref.~\cite{Ivan} as a function of $\beta f \Delta_b$.  The agreement
with (\ref{rateofhopping}) is excellent for $\beta f \Delta_b > 3.5$,
that is for $\phi > 0.6$, while at densities closer to $\phi_d$ the
agreement is less satisfactory.  But it should be noted that the
transition-state approximation is only expected to be good under the
same set of circumstances that make our approximations for the
equation of state good, that is, for $\phi \to \phi_{\text{max}}$.

\begin{figure}
\begin{center}
\includegraphics[width = 3.5in]{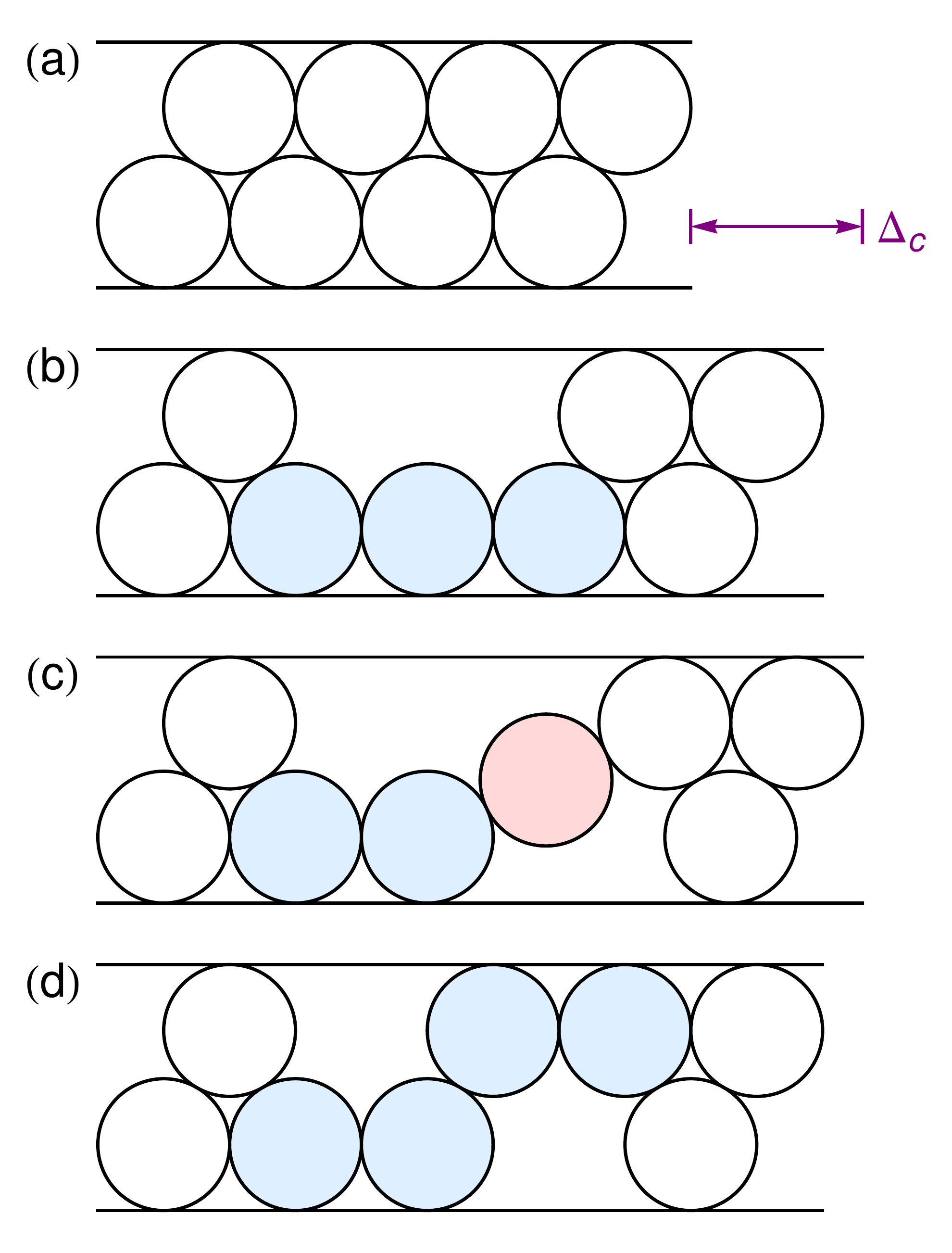}
\caption{(Color online) The creation of a pair of defects, starting
  from the most densely packed state~(a).  The intermediate state
  shown in (b) is unstable, as the middle of the three disks in a line
  can escape upwards.  Squeezing through the transition state shown in
  (c) leads to the two-defect state of (d).  $\Delta_c$ is the extra
  length occupied by the transition state. }
\label{twodefecttransitionstate}
\end{center}
\end{figure}

We next calculate the transition rate associated with the creation of
a pair of defects.  (With periodic boundary conditions, as used in
Ref.~\cite{Ivan}, defects can be created only in pairs.)  By detailed
balance, this rate of creation must equal the rate at which defects
move together and annihilate.  The transition state for creating a
pair of defects is shown in Fig.~\ref{twodefecttransitionstate}(c).
The extra length $\Delta_c$ required to reach this transition state is
\begin{equation}
\Delta_c=\sqrt{4 \sigma^2-h^2} +\sigma-3 \sqrt{\sigma^2-h^2}.
\label{twodefect}
\end{equation}
The transition rate for nucleating two defects or for the rate at
which pairs of defects annihilate is then
\begin{equation}
1/\tau_D=1/\tau_0 \exp(-\beta f \Delta_c).
\label{twodefectrate}
\end{equation}

The motion of the defects towards each other so that they might
annihilate is probably diffusive.  They typically have to diffuse a
distance of the order of their spacing $\xi$ to meet, so that one
might expect that
\begin{equation}
\tau_D/\tau \sim \xi^2,
\label{diffusion}
\end{equation}
where $\tau$ is the time scale for a defect to hop to a neighboring
site by the process illustrated in Fig.~\ref{transitionstate}.  Our
results for $\xi$, $\tau$ and $\tau_D$, are consistent with
Eq.~(\ref{diffusion}).  Note that this implies that the diffusion
coefficient for defects varies as~$1/\tau$.

Unfortunately there seem to be no direct studies of $\tau_D$ in the
molecular dynamics literature.  We can, however, use our expression
for $\tau_D$ to understand the results of the simulations described in
Ref.~\cite{Ashwin}.

\begin{figure}
\begin{center}
\includegraphics[width = 3.5in]{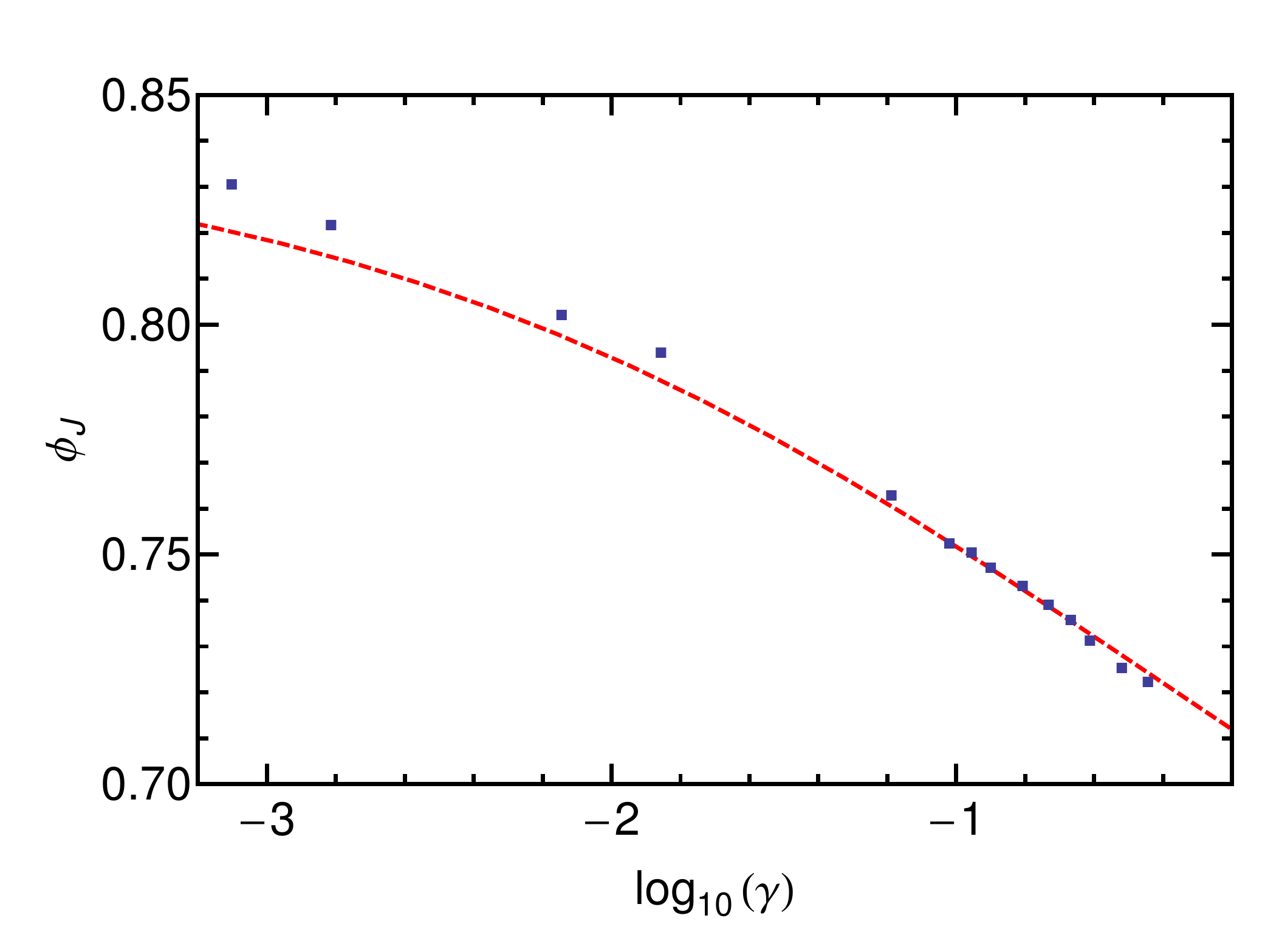}
\caption{(Color online) The jammed-state packing fraction $\phi_J$
  plotted against $\log_{10} (\gamma)$, where $\gamma$ is the quench
  rate.  The data points are taken from Ref.~\cite{Ashwin}.  The red
  dashed line is our prediction of $\phi_J(\gamma)$ from the
  Kibble--Zurek mechanism: see the discussion following
  Eq.~(\ref{phijtheta}).}
\label{phiJ}
\end{center}
\end{figure}

In this paper, the authors applied the Lubachevsky--Stillinger (LS)
algorithm \cite{LS} in which the diameter of the disks was increased
at a rate $\gamma =\sigma^{-1}\, d\sigma/dt$ in the course of a
molecular dynamics simulation starting from a small initial value
of~$\sigma$.  They kept the ratio $H_d/\sigma$ fixed and investigated
the $\gamma$-dependence of the jammed packing fraction $\phi_J$.
Their results for $\phi_J(\gamma)$ (shown in Fig.~\ref{phiJ}) indicate
that $\phi_J$ is a decreasing function of $\gamma$.
To understand this, it should first be noted that jammed states with
smaller values of $\phi_J$ contain more defects, the relationship
being~\cite{Mahdi}
\begin{align}
\phi_J&=\frac{N \pi \sigma^2}{4H_d[M \sigma+(N-M)\sqrt{\sigma^2-h^2}]} \nonumber\\
&=\frac{\pi \sigma^2}{4H_d[\theta \sigma+(1-\theta)\sqrt{\sigma^2-h^2}]}.
\label{phijtheta}
\end{align}
The hypothesis we shall make, following the ideas of Kibble
\cite{Kibble} and Zurek \cite{Zurek}, is that when the rate of
compression $\gamma$ exceeds the rate at which defects can annihilate
$1/\tau_D$, the defect density $\theta$ is frozen in and is not
changed in the last stages of the LS procedure.  Thus by equating
$1/\tau_D$ to $\gamma$ we get an estimate of $\beta f$ from
Eq.~(\ref{twodefectrate}).  This in turn can be used in Eq.~(\ref{A})
to obtain $\theta$, which then leads via Eq.~(\ref{phijtheta}) to
$\phi_J$.  These estimates of $\phi_J$ as a function of $\gamma$ are
plotted in Fig.~\ref{phiJ}, together with the simulational data of
Ref.~\cite{Ashwin}; there is reasonably good agreement between the
two.

Eq.~(\ref{phijtheta}) also shows that $\phi_J$ can be regarded as a
function of $\phi$, which we shall call $\phi_J(\phi)$, via the
dependence of $\theta$ on the equilibrium value of $\phi$.  We can use
$\phi_J(\phi)$ to rewrite Eq.~(\ref{equationB}) as
\begin{equation}
\beta f=\frac{2N}{L[1-\phi/\phi_J(\phi)]}.
\label{Stillingermap}
\end{equation}
Eq.~(\ref{rateofhopping}) expresses a relaxation time $\tau$ in terms
of $\beta f$ and so provides, in conjunction with
Eq.~(\ref{Stillingermap}), an illustration of the Stillinger map idea
that was used in Ref.~\cite{Barnett} to explain the relaxation times
of systems of hard disks and hard spheres.  The Stillinger map of a
configuration at packing fraction $\phi$ is its nearest inherent
(jammed) state, with packing fraction $\phi_J(\phi)$.  In the
narrow-channel system, the nearest jammed state is what is obtained in
an extremely rapid compression~\cite{Ashwin}.  Notice that for this
system $\beta f$ (and hence $\tau$) diverges only at $\phi
=\phi_{\text{max}}$: it is only at $\phi=\phi_{\text{max}}$ that
$\phi=\phi_J(\phi)$.

\section{Point-to-Set length}
\label{point2set}
In recent years it has been argued that the point-to-set length scale
$\xi_{PS}$ is an important length scale in glasses \cite{Cavagna,
  Cavagna2012}.  It is determined in, say, three dimensions by first
equilibrating the system of particles and then freezing all those
lying outside a spherical cavity of radius $R$.  One then studies the
correlation function $C(t) =\langle n(t) n(0)\rangle$ as $t \to
\infty$, where $n$ is the number of particles in a small box at the
center of the sphere.  If the cavity radius $R$ is greater than
$\xi_{PS}$, then $C(t)$ decays to the value it would have if the
particles moved randomly over the volume of the cavity.  However, when
$\xi_{PS} > R$ this does not happen, and so by varying $R$ one can
estimate $\xi_{PS}$.  Basically $\xi_{PS}$ is a measure of the size of
the smallest cavity for which the particles can escape their initial
positions when the particles on the boundary are frozen.

In our narrow-channel system we can mimic this procedure by simply
freezing all but $P$ disks, with the $P$ disks all adjacent to each
other.  Because in our system each disk interacts only with its
nearest-neighbors, one only needs to fix the two disks at the ends of
the region which contains the $P$ disks in typical configurations
drawn from an equilibrium distribution.  If the packing fraction is
$\phi$, then the length of the region will be $R= P\pi \sigma^2/(4H_d
\phi)$.  Provided $R-P\sqrt{\sigma^2-h^2} >\Delta_c$, the \lq \lq
cavity" will be large enough to allow the creation of the defects
which are needed to enable the system to relax.  This fixes a lower
bound on~$P$.  Setting $\xi_{PS}=P \sqrt{\sigma^2-h^2}$ gives
\begin{equation}
\xi_{PS} \sim \frac{\Delta_c \phi}{\phi_{\text{max}}-\phi}.
\label{PSlength}
\end{equation}
Note that $\xi_{PS}$  grows as a simple power law as $\phi \to
\phi_{\text{max}}$, whereas the correlation length $\xi$ grows
exponentially rapidly as $\phi \to \phi_{\text{max}}$, as can be seen
from Eqs.~(\ref{largexilimit}) and (\ref{SWlimit}).  As we have
pointed out earlier, $\xi$ (or rather $\xi\sqrt{\sigma^2-h^2}\,$) is
also the typical distance between defects.  Thus a region of size
$\xi_{PS}$ is unlikely to contain any defects in equilibrium, which
justifies the use of the relation $\xi_{PS}=P \sqrt{\sigma^2-h^2}$.
It also justifies the use of $\Delta_c$ (rather than $\Delta_b$) as
the additional length needed for relaxation, as there will be no
defects present to disrupt the zigzag order by their motion.

Activated dynamics is associated with the point-to-set length via
\begin{equation}
\tau=\tau_0 \exp[(\xi_{PS}/l)^{\psi}],
\end{equation}
where $l$ is a length scale of order $\sigma$.  For agreement with the
expressions for, say, $\tau_D$ or the $\tau$ of
Eq.~(\ref{rateofhopping}), we would require the exponent to be
$\psi=1$.  

The two length scales $\xi$ and $\xi_{PS}$ relate to different
phenomena.  In the glass literature $\xi_{PS}$ is popular
\cite{Cavagna, Cavagna2012}, as it does not require the identification
of the growing structural features which must be behind the onset of
caging.  In three dimensions, identifying the important clusters is
difficult and they may be dependent on details of the interatomic
potentials~\cite{Royall, Tanaka}.  But a full treatment of glass
behavior without such an understanding may be impossible.

\section{Discussion}
\label{discussion}
All of the equilibrium properties of the model of hard disks in a
channel can in principle be determined from the transfer matrix
integral equation, but solutions of this equation can only be obtained
numerically.  One of the purposes of our paper was to show the utility
of the analytical approximations which can be found for the limit when
$\phi \to \phi_{\text{max}}$.  In the same limit, the dynamics of the
system is essentially that of a dilute gas of defects.  We have
discussed some of its basic features, such as the time scales for the
creation and annihilation of defects and their diffusion rate.

As is often the case with exactly soluble models, it is hard to
calculate some particular quantities.  For example, the structure
factor $S(q_x, q_y)$ is the natural function to study for discussing
the growth of (say) zigzag order, but we do not know how to obtain it.
The dynamics of the model is non-trivial, and here much remains to be
done, via simulations and analytical approaches.  We have in mind here
the study and understanding of the autocorrelation function of
Eq.~(\ref{auto}).

One of the pleasing features of this model is that it has features
that mimic the behavior of three-dimensional spheres at the packing
fraction $\phi\simeq 0.58$; that is, it has an avoided dynamical
transition at $\phi_d$.  Above this density the dynamics involves
cooperative movements of the disks and is activated.  Approach to
$\phi_d$ from lower densities is acccompanied by the growth of the
length scale $\xi$ and when this is large enough, caging appears.  We
suspect that such a feature might be present in higher dimensions, but
the growth of the cage may require study of more subtle correlations
than those captured by the structure factor, for example those studied in
Ref.~\cite{Royall}.

\begin{acknowledgments}
We should like to thank Richard Bowles and Mahdi Zaeifi Yamchi for
sharing their data and insights.
\end{acknowledgments}


\begin{thebibliography}{99}

\bibitem{LiuNagel} A. J. Liu and S. R. Nagel, Nature (London), {\bf 396}, 21, (1998).

\bibitem{Mahdi} M. Z. Yamchi, S. S. Ashwin, and R. K. Bowles, Phys. Rev. Lett. {\bf 109}, 225701 (2012)

\bibitem{Ashwin} S. S. Ashwin, M. Zaeifi Yamchi, and R. K. Bowles, Phys. Rev. Lett. {\bf 110}, 145701 (2013).

\bibitem{Ivan} R. K. Bowles and I. Saika-Voivod, Phys. Rev. E {\bf 73}, 011503 (2006).

\bibitem{LS} B. D. Lubachevsky and F. H. Stillinger, J. Stat. Phys. {\bf 60}, 561 (1990).

\bibitem{Kibble} T. W. B. Kibble, J. Phys. A {\bf 9}, 1397 (1976).

\bibitem{Zurek} W. H. Zurek, Nature (London) {\bf 317}, 505 (1985).

\bibitem{Kofke} D. A. Kofke and A. J. Post, J. Chem. Phys. {\bf 98}, 4853 (1993).

\bibitem{Tonks} L. Tonks, Phys. Rev. {\bf 50}, 955 (1936).

\bibitem{Godfrey} M. J. Godfrey and M. A. Moore, to be published.

\bibitem{Krauth} E. P. Bernard and W. Krauth, Phys. Rev. Lett. {\bf 107}, 155704 (2011).

\bibitem{Kobdyn} W. Kob, S. Rold\'{a}n-Vargas, and L. Berthier, Nature Physics {\bf 8}, 164 (2012).

\bibitem{GodfreyB} M. J. Godfrey and M. A. Moore, unpublished.

\bibitem{Keys} A. S. Keys, A. R. Abate, S. C. Glotzer, and D. J. Durian, Nature Physics {\bf 3}, 260 (2007).

\bibitem{Narumi} T. Narumi, S. V. Franklin, K. W. Desmond, M. Tokuyama, and E. R. Weeks, Soft Matter {\bf 7}, 1472 (2011).

\bibitem{Brambilla} G. Brambilla, D. El Masri, M. Pierno, L. Berthier, L. Cipelletti, G. Petekidis, and A. B. Schofield, Phys. Rev. Lett., {\bf 102}, 085703 (2009).

\bibitem{SW} Z. W. Salsburg and W. W. Wood, J. Chem. Phys. {\bf 37}, 798 (1962).

\bibitem{Royall} A. Malins, J. Eggers, H. Tanaka, and C. P. Royall,
  Faraday Discuss. {\bf 167}, 405 (2013).

\bibitem{Tanaka} M. Leomach and H. Tanaka, Nature Communications {\bf 3}, 974 (2012).

\bibitem{Barnett} M. Barnett-Jones, P. A. Dickinson, M. J. Godfrey, T. Grundy, and M. A. Moore, Phys. Rev. E {\bf 88}, 052132 (2013).

\bibitem{Hunter} G. L. Hunter and E. R. Weeks, Phys. Rev. E {\bf 85}, 031504 (2012).

\bibitem{Cavagna}  C. Cammarota, A. Cavagna, G. Gradenigo, T. S. Grigera and P. Verrocchio, J. Chem. Phys. {\bf 131}, 194901 (2009); J. Stat. Mech. L12002 (2009).

\bibitem{Cavagna2012} A. Cavagna, T. S. Grigera, and P. Verrocchio, J. Chem. Phys. {\bf 136}, 204502 (2012).

\end{thebibliography}
\end{document}